\documentclass[a4paper,11pt]{article}
\usepackage{pos}

\usepackage{booktabs,array}
\usepackage[utf8]{inputenc}
\usepackage{float,graphicx,hhline}
\usepackage{algorithm,algpseudocode}
\usepackage{appendix}
\usepackage{mathtools}
\usepackage{multirow}
\usepackage{tikz-cd}
\usepackage{slashed}
\usepackage{soul}
\usepackage{pifont}
\usepackage{tikzducks}
\usepackage[caption=false]{subfig}
\usepackage[export]{adjustbox}
\usepackage{enumitem} %
\usepackage{bbm}
\usepackage{dsfont}
\usepackage{hyphenat}
\usepackage{comment}
\AtBeginDocument{
\heavyrulewidth=.08em
\lightrulewidth=.05em
\cmidrulewidth=.03em
\belowrulesep=.65ex
\belowbottomsep=0pt
\aboverulesep=.4ex
\abovetopsep=0pt
\cmidrulesep=\doublerulesep
\cmidrulekern=.5em
\defaultaddspace=.5em
}

\usepackage{hyperref}%
\hypersetup{
    pdftitle={},
    colorlinks=true,     %
    linkcolor=blue,      %
    citecolor=blue,      %
    filecolor=blue,      %
    urlcolor=blue        %
}
\usepackage[capitalize]{cleveref}
\crefname{equation}{Eq.}{Eqs.}
\crefname{section}{Sec.}{Secs.}
\Crefname{equation}{Equation}{Equations}

\DeclareMathAlphabet{\mathbbold}{U}{bbold}{m}{n}

\newcommand{\KL}{\mathrm{KL}}
\newcommand{\SU}{\mathrm{SU}}
\newcommand{\R}{\mathbb{R}}

\DeclareMathOperator{\Tr}{Tr}

\title{Multiscale Normalizing Flows for Gauge Theories}

\author*[a,b]{Ryan~Abbott}
\author[c]{Michael~S.~Albergo}
\author[a,b]{Denis~Boyda}
\author[f,a,b]{Daniel~C.~Hackett}
\author[a,b,d]{Gurtej~Kanwar}
\author[a,b]{Fernando~Romero-L\'opez}
\author[a,b]{Phiala~E.~Shanahan}
\author[a,b,e]{Julian~M.~Urban}

\affiliation[a]{Center for Theoretical Physics, Massachusetts Institute of Technology, Cambridge, MA 02139, USA}
\affiliation[b]{The NSF AI Institute for Artificial Intelligence and Fundamental Interactions}
\affiliation[c]{Center for Cosmology and Particle Physics, New York University, New York, NY 10003, USA}
\affiliation[d]{Albert Einstein Center, Institute for Theoretical Physics, University of Bern, 3012 Bern, Switzerland}
\affiliation[e]{Institut f\"ur Theoretische Physik, Universit\"at Heidelberg, Philosophenweg 16, 69120 Heidelberg, Germany}
\affiliation[f]{Fermi National Accelerator Laboratory, Batavia, IL 60510, U.S.A.}

\emailAdd{rabbott@mit.edu}

\abstract{
  Scale separation is an important physical principle that has
  previously enabled algorithmic advances such as multigrid solvers. Previous work on normalizing flows has been able to utilize scale separation in the context of scalar field theories, but the principle has been largely unexploited in the context of gauge theories. This work gives an overview of a new method for generating gauge fields using hierarchical normalizing flow models. This method builds gauge fields from the outside in, allowing different parts of the model to focus on different scales of the problem. Numerical results are presented for $U(1)$ and $SU(3)$ gauge theories in 2, 3, and 4 spacetime dimensions.

}

\FullConference{The 40th International Symposium on Lattice Field Theory (Lattice 2023)\\
July 31st - August 4th, 2023\\
Fermi National Accelerator Laboratory\\}

\begin{document}
\maketitle

\section{Introduction}

Normalizing flows~\cite{rezende2016variational,dinh2017density,JMLR:v22:19-1028} are a novel machine-learning based tool for sampling which has shown promise in the context of lattice field theory for alleviating or eliminating problems such as critical slowing down and
topological freezing~\cite{Alles:1996vn,DelDebbio:2004xh,Schaefer:2010hu}.  Applications of normalizing flows to lattice
field theories have seen great progress in recent years~\cite{Boyda:2022nmh, Cranmer:2023xbe}, with
demonstrations involving Abelian and non-Abelian gauge theories in 2,
3, and 4 dimensions, with and without fermions, including preliminary
work on QCD~\cite{Abbott:2022hkm}. %
In addition recent efforts in scaling from demonstrations in toy lattice volumes towards physically relevant theories have seen some success~\cite{fh,drrex}.

One physical principle which has yet to be fully integrated into
normalizing flow architectures for field theories is that of \emph{scale separation}.
Scale separation is the principle that physical processes and systems
can often be decomposed into separate processes occurring at
differing energy and length scales. Properly utilizing scale
separation is key element of calculations in many physical domains and in the context of lattice field theory has
lead to the development of many useful algorithms such as multigrid methods~\cite{Brannick:2007ue,Babich:2010qb}.
Previous machine-learning based work has explored the use of scale
separation in scalar field theories~\cite{Marchand:2022iwc,Hu:2020fbc,yu2020wavelet,guth2022wavelet}%
, 2D $U(1)$ gauge theory~\cite{Matsumoto:2023akw},  and in the context of linear preconditioners
\cite{Lehner:2023bba,Lehner:2023prf}, but not in the context of
non-Abelian gauge theories.

This work provides a construction of a new class of
hierarchically-constructed models referred to as \emph{multiscale
  models}. These models sample gauge fields by starting with coarse
degrees of freedom, and proceeding to add successively finer and finer
degrees of freedom. This structure allows these models to operate
separately on the UV and IR degrees of freedom, enabling more
expressive models that are able to take advantage of the physical
structure of the theory.

\section{Normalizing Flows}
Normalizing flows are a method from machine learning for
constructing an expressive, learned change of variables~\cite{rezende2016variational,dinh2017density,Albergo:2019eim}.
This change of variables takes the form of a parametrized bijective
map $f_{\theta}$ which can be used to transform a density $r(z)$
(typically referred to as the ``prior'' density) into a new ``model''
density $q_{\theta}(U)$ via
\begin{equation}
\label{eq:11}
q_{\theta}(U) = r(f^{-1}_{\theta}(U)) \left| \det \frac{\partial
	f^{-1}_{\theta}}{\partial U} \right|.
\end{equation}
Given a desired target density $p(U) = \frac{1}{Z} e^{-S(U)}$, the
model density can be trained to replicate the target density by
minimizing the reverse Kullback-Leibler (KL) divergence,\footnote{Note that the $\log Z$
term in (\cref{eq:kl-def}) is a constant independent of $U$, and hence
determining this (typically intractable) term is not required for
training.}
\begin{align}
\label{eq:3}
\KL(q || p) &= \mathbb{E}_{U \sim q(U)}[\log q(U) - \log p(U)] \\
\label{eq:kl-def} &= \mathbb{E}_{U \sim q(U)}[\log q(U) + S(U)] + \log Z ~. %
\end{align}

Once a model has been trained, the target density $p(U)$ can be
sampled via methods such as independence Metropolis~\cite{Albergo:2019eim,noe2019boltzmann,Metropolis:1953am,Hastings:1970aa,tierney1994markov}, or direct
reweighting from the model density. The efficiency of such methods can
be estimated using the effective sample size (ESS) statistic, defined
by~\cite{doucet2001sequential,liu2001monte}
\begin{equation}
  \label{eq:7}
  \mathrm{ESS} = \frac{\mathbb{E}_{U \sim q(U)}[w(U)]^2}{\mathbb{E}_{U
	\sim q(U)}[w(U)^2]},
\end{equation}
where $w(U) = \frac{e^{-S(U)}}{q(U)}$ is the (unnormalized)
reweighting factor associated to $U$. The ESS gives an approximation
of how many independent target samples are obtained per model sample,
and the ESS is always bounded between 0 and 1.

\section{Multiscale Models in Two Dimensions}

\begin{figure}[t]
\centerline{\includegraphics[]{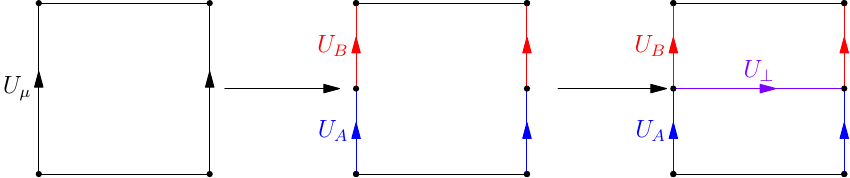}}
\caption[]{\label{fig:doubling}Schematic of the procedure for doubling
layers, which serve to double the number of lattice sites in a particular spacetime direction} %
\end{figure}

Similar to a typical normalizing flow model, multiscale models begin
by sampling from a prior density, referred to as the \emph{coarse
prior}. Unlike most normalizing flow-based models, however, samples
from the coarse prior do \emph{not} live in the same space as target
samples; instead the samples from the coarse prior are gauge fields
living on a coarser lattice, possibly as coarse as a single lattice
site. The coarse prior can be any tractable density that can be
sampled; for instance, the coarse prior might be taken as a gauge
theory with a coarser lattice spacing than the target density, or the
coarse gauge links could be sampled from the uniform (Haar) distribution.

Once the coarse degrees of freedom have been sampled, new degrees of freedom
are added via successive \emph{doubling layers}. A doubling layer
takes as input a set of ``coarse'' gauge links
$U_{\nu}^{\text{coarse}}$ living on a lattice $\Lambda$ along with a
\emph{doubling direction} $\mu$, and outputs a
new set of ``fine'' gauge links $U_{\nu}^{\text{fine}}$ arranged on a lattice $\Lambda'$ with
twice as many sites appearing along the $\hat{\mu}$
direction. Conceptually, the lattice $\Lambda'$ is considered to be a
refinement of $\Lambda$, meaning that $\Lambda$ and $\Lambda'$ are
both considered subsets of the same underlying space with $\Lambda
\subset \Lambda'$. This means that applying a doubling layer halves
the lattice spacing in the $\hat{\mu}$-direction, leading to an
anisotropic lattice spacing $a_\nu$.  By applying successive doubling layers
with different doubling directions, the underlying lattice spacing can
be made arbitrarily small in every direction. 

Doubling layers can be constructed in arbitrary dimensions; however, it is simpler to first construct 2-dimensional doubling layers, and then later generalize the layers to accommodate arbitrary dimensions (see \cref{sec:arbitrary-dim}). A 2-dimensional doubling layer with doubling direction $\mu$ can be
separated into two steps, as illustrated in
\cref{fig:doubling}. First each gauge link oriented along the
$\hat{\mu}$-direction is split into two gauge links $U_A$ and $U_B$,
subject to the condition
\begin{equation}
\label{eq:4}
U_{\mu}(x) = U_A(x) U_B(x + a_\mu \hat{\mu} /2) ~.
\end{equation}
This can be achieved by sampling $U_B$ from the uniform (Haar)
distribution on the given gauge group, and then defining $U_A$ from \cref{eq:4}.
Note that at this point no new physical information has been added,
since the added gauge links $U_A$ and $U_B$ do not create any new
loops. All physical information is added in the second part of the
doubling layer, wherein a new set of perpendicular links $U_{\perp}$
is sampled from a chosen distribution. For instance, one possible
distribution for sampling $U_{\perp}$ would be the heatbath-like
distribution
\begin{equation}
\label{eq:5}
p_{HB}(U_{\perp}) \propto \exp \left\{ \tilde{\beta} \sum_x
  \mathrm{Re} \Tr [
  U_{\perp}(x + a_\mu \hat{\mu} / 2) (S_A + S_B)
(x + a_\mu \hat{\mu} / 2)]
\right\}
\end{equation}
where $\tilde{\beta} \in \R$ parameterizes the probability
distribution, and $S_A$ and $S_B$ are the staples defined by
\begin{align}
\label{eq:SA-def} S_A(x + a_\mu \hat{\mu} / 2)
  &=U_A^{\dag}(x + a_\nu \hat{\nu}) U_{\nu}^{\dag}(x) U_A(x) \\
\label{eq:SB-def} S_B(x + a_\mu \hat{\mu} / 2)
  &= U_B(x + a_\mu\hat{\mu}/2 + a_\nu \hat{\nu}) U_{\nu}^{\dag}(x + a_\mu \hat{\mu})
	U_B^\dag(x + a_\mu\hat{\mu}/2)
\end{align}
and $\nu$ is the direction orthogonal to $\mu$. Alternatively
$U_{\perp}$ could be a sampled from a normalizing flow model conditioned
on $S_A$ and $S_B$, as will be discussed in \cref{sec:staple-conditional}.

Once $U_{\perp}$ has been generated, the final fine lattice
$U_{\nu}^{\text{fine}}$ can be assembled via
\begin{equation}
\label{eq:8}
U_{\nu}^{\text{fine}}(x) =
\begin{cases}
U_{\nu}^{\text{coarse}}(x) & x_{\mu} / a_\mu' \text{ even and } \nu \neq \mu \\
U_{\perp}(x) & x_{\mu} / a_\mu' \text{ odd and } \nu \neq \mu \\
U_{A}(x) & x_{\mu} / a_\mu' \text{ even and } \nu = \mu \\
U_{B}(x) & x_{\mu} / a_\mu' \text{ odd and } \nu = \mu \\
\end{cases}
\end{equation}
where $a_\mu' = a_\mu / 2$ indicates the lattice spacing of the fine lattice $\Lambda'$ in the direction $\mu$.
This completes the construction of the doubling layers in 2 dimensions; all that remains is to compute the density of the resulting field $U_\nu^{\text{fine}}$, which can be accomplished using the factorization
\begin{equation}
\label{eq:doubling-layer-density-factorization}
    q(U_\nu^\text{fine}) = q(U_\perp \mid U_A, U_B, U_\nu^\text{coarse}) q(U_B \mid U_\nu^\text{coarse})
    q(U_\nu^\text{coarse}).
\end{equation}
Here $q(U_\perp \mid U_A, U_B, U_\nu^\text{coarse})$ is the model density of a staple-conditional model, which will be discussed in \cref{sec:staple-conditional}, and $U_B$ is sampled from the Haar distribution, which has density $q(U_B \mid U_\nu^\text{coarse}) = 1$.
Note here that $U_A$ is determined by $U_B$ and $U_\nu^{\text{coarse}}$, and hence does not require an additional term in \cref{eq:doubling-layer-density-factorization}.

\subsection{Staple-Conditional Models}\label{sec:staple-conditional}

In order to sample $U_{\perp}$ (see \cref{fig:doubling}), it is desirable to have a class of
expressive, conditional models that can incorporate as much
gauge-equivariant information from the local neighborhood of
$U_{\perp}$ as possible.
One natural piece of gauge-equivariant conditional information are the
staples $S_A$ and $S_B$ defined in \cref{eq:SA-def,eq:SB-def}. More
generally, any composition of gauge links passing from the endpoint to
the starting point of $U_{\perp}$ has the same gauge transformation
properties as $S_A$ and $S_B$, and hence can be considered as an
abstract staple. Attempting to integrate this information into the
sampling of $U_{\perp}$ then leads to the general notion of
staple-conditional models, which are models capable of sampling a single
gauge matrix $U$, conditioned on a particular number of
``staples'' $S_1, \dots, S_n$. Note that this architecture requires that each gauge link within $U_\perp$ is sampled independently conditioned on the staples. Staple-conditional models can be based
on any tractable distribution; in this work we will primarily consider
staple-conditional models based on normalizing flows.

There are two components to a normalizing flow-based
staple-conditional model: a prior (or base) distribution $r(z \mid
S_1, \dots, S_n)$, $z \in G$, and a conditional normalizing flow $f(z
\mid S_1, \dots, S_n)$. These two pieces can then be combined to
define a model density $q(U \mid S_1, \dots, S_n)$:
\begin{equation}
\label{eq:2}
q(U \mid S_1, \dots, S_n) = r(f^{-1}(U) \mid S_1, \dots, S_n)
\left| \det \frac{\partial f^{-1}(U)}{\partial U} \right| ~.
\end{equation}
The prior distribution can be any tractable distribution, such as the
uniform (Haar) distribution, or the heatbath-type distribution defined
by \cref{eq:5} in the case of a $U(1)$ gauge theory.\footnote{Note
that here ``tractable'' includes the requirement of computing the
normalized density, which is not possible in the case of a $\SU(3)$
heatbath due to the lack of an analytic formula for the normalizing
constant for the distribution \cref{eq:5}. Computing the normalizing constant as well as its gradients numerically is theoretically possible, but the computational cost of doing so would be impractically large for the models considered in \cref{sec:2d-results,sec:4d-results}.} For the conditional
normalizing flow, the tools and methods developed for constructing
gauge-equivariant flow models can be adapted to the problem with
slight modifications. For instance, in the $\SU(N)$ case both spectral
and residual flows~\cite{Abbott:2023thq} provide building blocks that can be
composed to create expressive conditional transformations. For the
$\SU(N)$ staple-conditional flows involved in this work, a spectral
flow is used, acting on an effective active loop $P$ defined by
\begin{equation}
\label{eq:9}
P = \mathrm{Proj}_{\SU(N)} \left\{ U \sum_{i = 1}^n \alpha_i S_i \right\}
\end{equation}
where $\{\alpha_i\}_{i=1}^n$ are learned coefficients, and
$\mathrm{Proj}_{\SU(N)}$ is a projection onto $\SU(N)$, accomplished
via polar decomposition:
\begin{equation}
\label{eq:10}
\mathrm{Proj}_{\SU(N)}(M) = \frac{M (M^{\dag} M)^{-1/2}}{\det \left[ M
  (M^{\dag} M)^{-1/2}\right]^{1/N}}.
\end{equation}
The spectral flow acting on $P$ produces a new, transformed value $P'
\in \SU(N)$. This transformation can then be pushed back to the gauge
matrix $U$ to produce a new matrix $U'$ via
\begin{equation}
U' = P' P^{\dag} U.
\end{equation}
This procedure can then be iterated with different values of
$\alpha_i$ and different spectral flows in order to construct an
expressive, learnable transformation. Alternatively, architectures based on residual flows~\cite{Abbott:2023thq} may also be used as coupling layers, either in place of or alongside spectral flows. Although the numerical demonstrations in \cref{sec:2d-results,sec:4d-results} utilize only spectral flow based staple conditional models, simiar models based on residual flows achieve similar performance.

For $U(1)$ gauge fields, staple-conditional models can be constructed in a similar manner as the $\SU(N)$ staple-conditional models. The numeric demonstrations in \cref{sec:2d-results} are constructed using
the heatbath-like prior defined in \cref{eq:5} combined with a sequence of coupling layers based on circular rational quadratic splines.~\cite{Kanwar:2020xzo} The coupling layers in the $U(1)$ case have the same structure as the $\SU(N)$ staple-conditional flows, except that the projection in \cref{eq:9} is replaced with a normalization step
\begin{equation}
    \mathrm{Proj}_{U(1)}(z) = \frac{z}{|z|}.
\end{equation}

\subsection{Numeric Results}\label{sec:2d-results}

\begin{figure}[t]
\centerline{\includegraphics[width=0.5\textwidth]{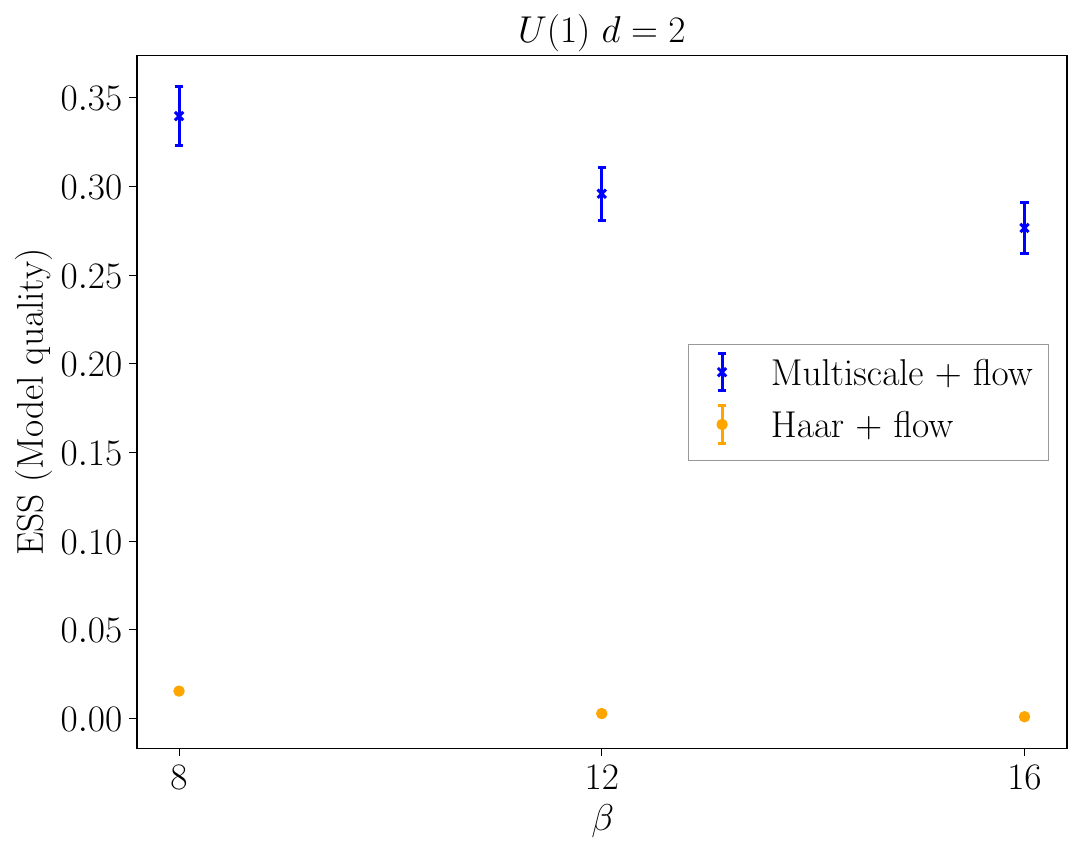}
  \hspace{0.5cm}
\includegraphics[width=0.5\textwidth]{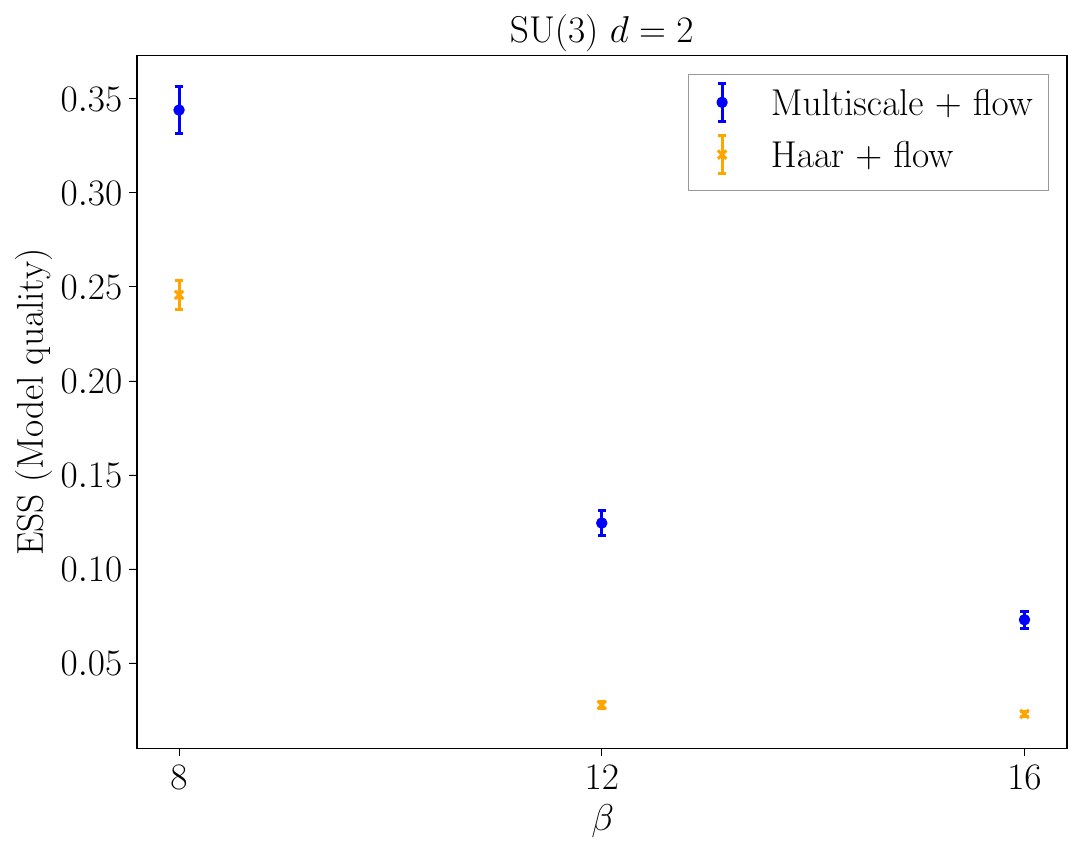}}
\caption[]{\label{fig:2d-results}Effective sample size (ESS) for 2-dimensional multiscale models (higher is better). Blue data points indicate the performance of the multiscale model combined with a small fine-lattice flow, while the orange data points represent the fine-lattice flow alone.}
\end{figure}

Numerical demonstrations of multiscale models are investigated for both $U(1)$ and $\SU(3)$ gauge theories in 2 dimensions; results are shown in \cref{fig:2d-results}.
All models start from a Haar
uniform coarse prior on a $2 \times 2$ lattice, and target an $8 \times
8$ fine lattice. In all cases the multiscale model has been
integrated into the overall model as a prior, with an additional flow
applied at the finest scale after all of the degrees of freedom have
been generated. This is necessary due to the fact that the multiscale
models as constructed only attempt to capture a subset of possible
correlations in the output gauge configurations, and hence a full
fine-lattice flow is still required for maximum expressivity.
For the $U(1)$ models the
fine lattice flow consists of 48 gauge-equivariant coupling layers
utilizing rational quadratic splines~\cite{Kanwar:2020xzo}, while the
find lattice flows for the $\SU(3)$ models consist of a single iteration of direction and
location updates, for a total of $8 + 4 = 12$ layers in 2 dimensions~\cite{Abbott:2023thq}.

In both the $U(1)$ and $\SU(3)$ cases the multiscale models show
significant improvement over equivalent models without the multiscale component used as a prior. In the $U(1)$ case the quality of the
multiscale models is effectively independent of $\beta$, while the
non-multiscale models have an effective sample size consistent with
0. Meanwhile in the $\SU(3)$ case the model quality does decline with
$\beta$, but more slowly for the multiscale models, which maintain
$\sim 10\%$ ESS even for $\beta = 16$.

\section{Multiscale models in higher dimensions}\label{sec:arbitrary-dim}

\begin{figure}[t]
\centerline{\includegraphics[width=0.8\textwidth]{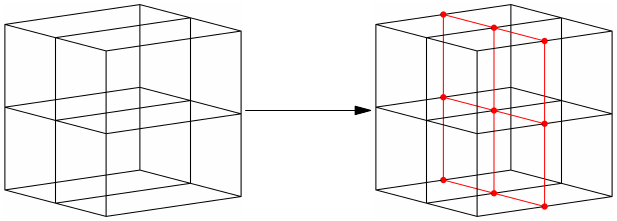}}
\caption[]{\label{fig:3d2dslice} Illustration of a 3-dimensional
  doubling layer; red points indicate the new lattice sites in the
  doubled lattice, and red lines indicate the added gauge links.}
\end{figure}

In greater than two dimensions, models can be constructed in the same
manner as in two dimensions, with a few additional complications. As
in two dimensions, higher-dimensional multiscale models are also
constructed as a sequence of doubling layers, each doubling the
lattice extent along a particular direction. The primary complication
in adapting the 2-dimensional models to higher dimensions occurs
during the generation of the new gauge links $U_{\perp}$ that are
orthogonal to the doubling direction. In $d$-dimensions, the role that
$U_{\perp}$ plays in 2-dimensional models is replaced by a $(d -
1)$-dimensional slice of the lattice, as illustrated in
\cref{fig:3d2dslice}. This slice contains both UV and IR degrees of
freedom, and hence more care is required handling this slice than in
the 2-dimensional case.

One viable solution for sampling the $(d - 1)$-dimensional slice
needed in a $d$-dimensional doubling layer is to utilize a $(d -
1)$-dimensional multiscale model. This gives the models a recursive
structure, with each multiscale model depending recursively on
several lower-dimensional models, terminating at the base case of a
2-dimensional model, for which the previously described 2-dimensional
multiscale models are sufficient. This gives the broad structure of
the model, which is subject to a handful of complications.

The first complication arises in the conditioning of the models. In
order to ensure that the lower dimensional models can generate correlations in the perpendicular field $U_\perp$, 
models need to have access to sufficient gauge-equivariant
information, including information not present within the slice that
the lower-dimensional model operates within. In order to pass such
information, the multsicale models as described above can be modified
by adding additional conditioning in the form of staples constructed
out of gauge links present in the higher-dimensional models.
This means that, for instance, a 2-dimensional doubling layer
contained within a 3-dimensional model will also receive as input
staples which reach outside of the 2-dimensional plane. These staples
can be passed to the staple-conditional models present within the
doubling layers, allowing the staple-conditional models to properly
account for out-of-plane information.

The second complication of the higher dimensional doubling layers
arises during the first part of the doubling layers, wherein the gauge
links oriented along the doubling direction are split in half (see
\cref{eq:5}). In 2-dimensional doubling layers this step does not add
any new physical information, but this is not necessarily true if the
2-dimensional doubling layer is a subcomponent of a higher-dimensional
model. In particular, if there already exist paths of links
connecting the two endpoints of $U_B$, then $U_B$ could be sampled
from a staple-conditional model conditioned on these paths, adding new
physical information to the sampling step. Furthermore, it is also possible to sample $U_B$ at some sites before others, allowing the later values of $U_B$ to be conditioned on previous values. In practice this requires
careful bookkeeping to route the appropriate staples into a new
staple-conditional model for sampling $U_B$.
\begin{algorithm}
\caption{Generic $d$-dimensional doubling layers}\label{alg:doubling-layers}
\begin{algorithmic}[1]
    \State \textbf{Input}: Coarse gauge field $U^\text{coarse}_\nu$, doubling direction $\mu$, higher-dimensional staples $S$
    \State Sample $U_B$ from staple-conditional model
    \State Compute $U_A = U_\mu U_B^\dag$
    \State Add $S_A, S_B$ from \cref{eq:SA-def,eq:SB-def} to $S$
    \If{$d = 2$}
    \State Sample $U_\perp$ from staple-conditional model
    \Else
    \State Sample $(d - 1)$-dimensional slices $U_\perp$ from $(d - 1)$-dimensional multiscale model
    \EndIf
    \State Combine $(U^\text{coarse}_\nu, U_A, U_B, U_\perp)$ via \cref{eq:8}
    \State \textbf{Output}: fine gauge field $U^\text{fine}_\nu$
\end{algorithmic}
\end{algorithm}

\begin{algorithm}
    \caption{$d$-dimensional multiscale model}
    \begin{algorithmic}
    \State \textbf{Input}: Higher-dimensional staples $S$
    \State Sample gauge field $U_\nu$ at coarsest scale
    \While{$U_{\nu}$ has not reached finest scale}
	\State Choose doubling direction $\mu$
    \State $U_\nu \gets \mathrm{DoublingLayer}(U_\nu, \mu, S)$
    \EndWhile
    \end{algorithmic}
\end{algorithm}

\subsection{Numeric Results}\label{sec:4d-results}
\begin{figure}[t]
\centerline{\includegraphics[width=0.5\textwidth]{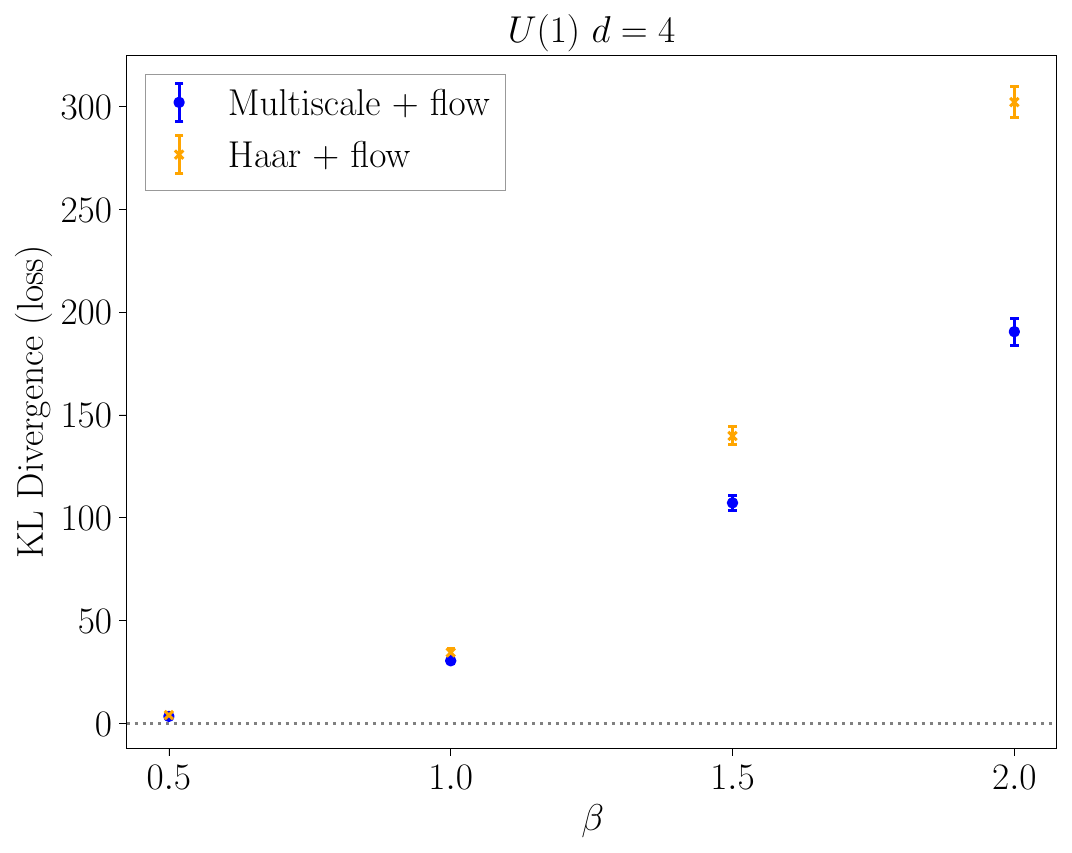}
  \hspace{0.5cm}
\includegraphics[width=0.5\textwidth]{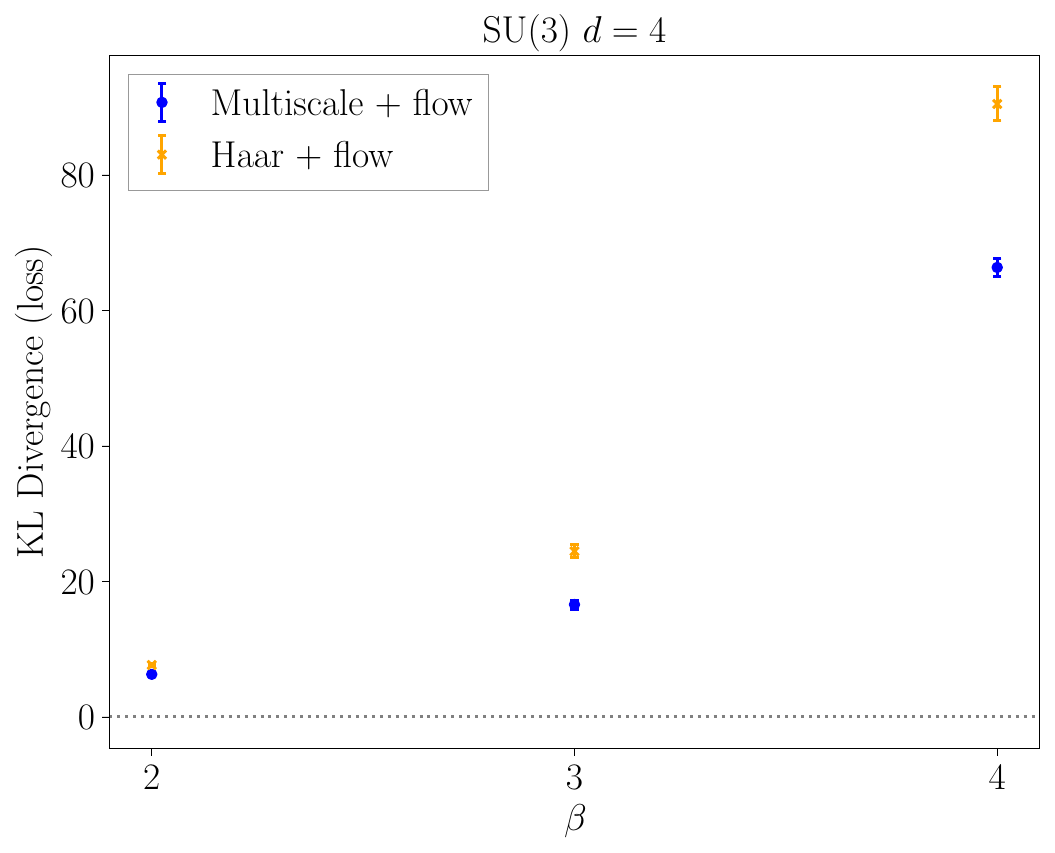}}
\caption[]{\label{fig:4d-results} KL divergence (see \cref{eq:kl-def})
  for 4-dimensional multiscale models (lower is better). Blue data
  points indicate the performance of the multiscale model combined
  with a small fine-lattice flow, while the orange data points
  represent the fine-lattice flow alone.
  }
\end{figure}

Numeric results for 4-dimensional models for both $U(1)$ and $\SU(3)$ gauge theories are shown in
\cref{fig:4d-results}, in terms of the KL divergence as defined in
\cref{eq:kl-def}. Both models start from a Haar uniform coarse prior
of size $L / a_{\text{coarse}} = 2$, and target a periodic fine lattice of size $L /
a_{\text{fine}} = 4$. The constant $\log Z$ in \cref{eq:kl-def} is estimated from the
best model available at the given
$\beta$, and the same value of $\log Z$ is used for every model at the
same value of $\beta$. As with the 2-dimensional models, the
multiscale models here are integrated into a larger model as a prior,
with a small additional fine-lattice flow. For the $U(1)$ models the
fine lattice flow consists of 48 gauge-equivariant coupling layers
utilizing rational quadratic splines \cite{Kanwar:2020xzo}, while the
fine lattice flows for the $\SU(3)$ models consist of a single iteration of direction and
location updates, for a total of $48 + 16 = 64$ layers in 4 dimensions. \cite{Abbott:2023thq}

In all cases the KL divergence of the multiscale models is smaller
than the KL divergence of the fine-lattice flow alone. The degree of
improvement is slight at smaller values of $\beta$, but increases for
larger values of $\beta$, indicting that the multiscale models have
increased expressivity at higher $\beta$ relative to the fine-lattice
flows utilized in this work. One plausible explanation for this
difference is that the higher $\beta$ distributions tend to have
stronger correlations at longer distances. Multiscale models can build
such correlations directly when acting at the coarser scales, whereas
the full fine-lattice flow needs to build in the correlation structure
starting from the finest scale upwards, which is particularly
difficult for the relatively small fine-lattice flows utilized in this
work.

\section{Future work}
There are several possible avenues for future improvements for the
models presented here.

\begin{itemize}
\item The fine-lattice flows utilized in the numeric results here are
intentionally small in order to isolate the effects of the multiscale
models. Future studies could combine the multiscale models with
larger, more expressive fine-lattice flows in order to obtain a
maximally expressive combination.
\item For the sake of simplicity, the multiscale models presented here
assume independence of links sampled at a given scale after
conditioning on the coarser scales. Future models could relax this
assumption, which would also have the benefit of making the multiscale
models universal density approximators, meaning that the models could
(in principle) approximate any density to an arbitrary degree of
precision, given a sufficient number of parameters.
\item The present work has focused on the method of direct sampling as
a benchmarking task for these new models; however, this is not
necessarily the most efficient method for utilizing multiscale
models. Instead a hybrid approach that combines more traditional MCMC
methods such as HMC with flow-based methods could provide a better
avenue for near-term physics applications (see, for instance, the
recent work in Refs.~\cite{fh,drrex}).
\end{itemize}

\section{Conclusion}
Multiscale models show great potential for improving near- and far-term
normalizing flow capabilities in the context of lattice field
theory. By operation on the UV and IR degrees of freedom separately,
these models are able to exploit scale separation in order to more
effectively replicate the desired gauge field distribution. Though
more work is needed to extend these models and integrate them more
fully with other methods, this represents a promising step in
understanding and implementing normalizing flows for gauge generation,
and more broadly improving the efficiency of gauge generation as a whole.

\section*{Acknowledgements}

We thank Aleksandar Botev, Kyle Cranmer, Alexander G.~D.~G.\ Matthews, S\'ebastien Racani\`ere, Ali Razavi, and Danilo J. Rezende for useful discussions and valuable contributions to the early stages of this work. RA, DCH, FRL, PES, and JMU are supported in part by the U.S.\ Department of Energy, Office of Science, Office of Nuclear Physics, under grant Contract Number DE-SC0011090. PES is additionally supported by the U.S.\ DOE Early Career Award DE-SC0021006, by a NEC research award, and by the Carl G and Shirley Sontheimer Research Fund. FRL acknowledges support by the Mauricio and Carlota Botton Fellowship. GK was supported by the Swiss National Science Foundation (SNSF) under grant 200020\_200424. This manuscript has been authored by Fermi Research Alliance, LLC under Contract No.~DE-AC02-07CH11359 with the U.S. Department of Energy, Office of Science, Office of High Energy Physics. This work is supported by the U.S.\ National Science Foundation under Cooperative Agreement PHY-2019786 (The NSF AI Institute for Artificial Intelligence and Fundamental Interactions, \url{http://iaifi.org/}) and is associated with an ALCF Aurora Early Science Program project, and used resources of the Argonne Leadership Computing Facility which is a DOE Office of Science User Facility supported under Contract DEAC02-06CH11357. The authors acknowledge the MIT SuperCloud and Lincoln Laboratory Supercomputing Center~\cite{reuther2018interactive} for providing HPC resources that have contributed to the research results reported within this paper. Numerical experiments and data analysis used PyTorch~\cite{NEURIPS2019_9015}, JAX~\cite{jax2018github}, Haiku~\cite{haiku2020github}, Horovod~\cite{sergeev2018horovod}, NumPy~\cite{harris2020array}, and SciPy~\cite{2020SciPy-NMeth}. Figures were produced using matplotlib~\cite{Hunter:2007}.

\bibliographystyle{JHEP}
\bibliography{main}

\end{document}